\documentclass{article}
\usepackage{spconf,amsmath,graphicx,url,hyperref}
\usepackage{color}
\usepackage{booktabs}

\usepackage{amsfonts}

\usepackage{lineno}
\usepackage{subcaption}

\usepackage[style=ieee,
uniquelist=false,
mincitenames=1,
maxcitenames=2,
minbibnames=1,
maxbibnames=2,
backend=biber]{biblatex}
\title{Bass Accompaniment Generation via Latent Diffusion}
\name{Marco Pasini \textsuperscript{1,2}, Maarten Grachten \textsuperscript{1}, Stefan Lattner \textsuperscript{1}}
\address{Sony Computer Science Laboratories, Paris, France\textsuperscript{1} \qquad Queen Mary University, London, UK\textsuperscript{2}}

\newcommand{\style}{\ensuremath{\mathit{style}}}
\newcommand{\ground}{\ensuremath{\mathit{ground}}}
\newcommand{\cfg}{\ensuremath{\mathit{cfg}}}
\newcommand{\dimension}{\ensuremath{\mathit{dim}}}
\newcommand{\rtime}{\ensuremath{r_\mathit{time}}}
\newcommand{\mssd}{\ensuremath{\mathit{mssd}}}
\newcommand{\hoplen}{\ensuremath{\mathit{hop\_len}}}
\newcommand{\winlen}{\ensuremath{\mathit{win\_len}}}
\newcommand{\lambdarec}{\ensuremath{\lambda_{\mathit{rec}}}}
\newcommand{\lambdamssd}{\ensuremath{\lambda_{\mssd}}}

\newcommand{\rec}{\ensuremath{\mathit{rec}}}
\newcommand{\adv}{\ensuremath{\mathit{adv}}}
\newcommand{\magnitude}{\ensuremath{\mathit{mag}}}
\newcommand{\STFT}{\ensuremath{\mathrm{STFT}}}
\newcommand{\iSTFT}{\ensuremath{\mathrm{iSTFT}}}

\begin{document}
%
\maketitle

\begin{abstract}
The ability to automatically generate music that appropriately matches an arbitrary input track is a challenging task.
We present a novel controllable system for generating single stems to accompany musical mixes of arbitrary length.
At the core of our method are audio autoencoders that efficiently compress audio waveform samples into invertible latent representations, and a conditional latent diffusion model that takes as input the latent encoding of a mix and generates the latent encoding of a corresponding stem.
To provide control over the timbre of generated samples, we introduce a technique to ground the latent space to a user-provided reference style during diffusion sampling.
For further improving audio quality, we adapt classifier-free guidance to avoid distortions at high guidance strengths when generating an unbounded latent space.
We train our model on a dataset of pairs of mixes and matching bass stems.
Quantitative experiments demonstrate that, given an input mix, the proposed system can generate basslines with user-specified timbres.
Our controllable conditional audio generation framework represents a significant step forward in creating generative AI tools to assist musicians in music production.
\end{abstract}
%
\section{Introduction}
\label{sec:intro}
Musical accompaniment is an integral part of music composition and performance. The ability to automatically generate an accompaniment that complements and matches the style of existing instrument parts (\emph{stems}) in a music track, has the potential to both enhance the creativity of artists--by proposing novel musical material for them to work with--and to make it easier and more efficient to realize their artistic visions.
In recent years, deep learning techniques have shown promising results in the field of music and (to a much lesser extent) accompaniment generation.
Many approaches use a symbolic representation of music as the medium \cite{deepbach,musictransformer,figaro}, while more recently a number of models that directly generate waveform audio have also been proposed \cite{jukebox,musiclm,gensep}.
Diffusion models \cite{diffusion,ddpm,ddim} have emerged as a powerful class of generative models capable of producing high-quality samples, although they usually require a computationally expensive iterative sampling procedure.
Latent diffusion models \cite{latdiff} have been introduced to increase model inference speed by generating a latent, low-dimensional representation of the data from a pretrained autoencoder model, usually a Variational AutoEncoder \cite{vae}.

In this work, we propose a general latent generative model for the task of accompaniment generation, and apply it to the generation of basslines.
Given an input stem of arbitrary length such as a vocal melody or an input mix of arbitrary numbers of stems, our model is able to generate a complementary bass stem that musically matches the conditioning.
Furthermore, we propose controllability features, such as style conditioning and conditioning guidance control, to make our system a more useful tool for artists.
The key contributions of our work are:

\begin{itemize}
\item The design of an efficient audio autoencoder to encode samples to compressed invertible representations
\item The design of a general conditional latent diffusion model that takes a music mix as input and produces a coherent track, while being able to handle inputs and outputs of arbitrary length
\item The application of both audio autoencoder and latent diffusion model to the task of encoding and generating basslines given an arbitrary input mix
\item The use of style conditioning during the diffusion sampling process to force the generation of a user-defined bass style.
\end{itemize}

\section{Related Work}

\label{sec:related_work}

Accompaniment generation is a type of music generation that involves an additional input conditioning.
In this work we focus on audio-based music generation.
Autoregressive models such as WaveNet \cite{wavenet}, SampleRNN \cite{samplernn},  Jukebox \cite{jukebox}, MusicLM \cite{musiclm} and MusicGen \cite{musicgen} can generate high quality samples but suffer from slow sequential sampling.
Non-autoregressive models based on generative adversarial networks (GANs) \cite{gan} such as WaveGAN \cite{wavegan} and GANSynth \cite{gansynth} achieve parallel sampling but are limited to generating fixed-length audio clips.
On the other hand, Musika \cite{musika} parallelly generates invertible latent representations of audio of arbitrary length, but the context available to the model is limited.
Relevant to our work, BassNet \cite{bassnet} generates bass tracks while offering user control via a latent space variable.

More recently, models such as DiffWave \cite{diffwave} and WaveGrad \cite{wavegrad} introduce diffusion to audio modeling for speech synthesis applications.
For musical audio generation, Riffusion \cite{riffusion} fine-tunes Stable Diffusion \cite{latdiff} on audio spectrograms to generate music clips.
Moûsai \cite{mousai} trains a latent diffusion model on compressed representations and can generate minute-long coherent music.
JEN-1 \cite{jen1} introduces a large-scale conditional latent diffusion model that can generate long-form music both autoregressively and non-autoregressively.
Finally, \cite{gensep} proposes a multi-source diffusion model trained on single source waveforms that achieves both generation and separation of individual sources. 



\section{Method}
\label{sec:method}
Let $\mathbf{x}=\{\mathit{x_1},...,\mathit{x_T}\}$ be the waveform of a mix of arbitrary stems of length $T$, where $x_i$ is the $i$-th stereo frame, and let $\mathbf{y}=\{\mathit{y_1},...,\mathit{y_T}\}$ be the waveform of a single-stem audio sample with the same length.
To sample $\mathbf{y}$ given $\mathbf{x}$, we aim to model the conditional distribution $\mathit{p(\mathbf{y}|\mathbf{x})}$, but since the waveforms are typically very high-dimensional (i.e.
$T$ is large), we encode both $\mathbf{x}$ and $\mathbf{y}$ into latent representations $\mathbf{c_x}=\{\mathit{c_{x,1}},...,\mathit{c_{x,T/\rtime}}\}$ and $\mathbf{c_y}=\{\mathit{c_{y,1}},...,\mathit{c_{y,T/\rtime}}\}$ respectively using audio autoencoders, and model $\mathit{p(\mathbf{c_y}|\mathbf{c_x})}$ instead.
Here, $\rtime$ is the time compression ratio of the autoencoders, and we refer to the dimensionality of vectors $\mathit{c_{x,i}}$ and $\mathit{c_{y,i}}$ as $\dimension_x$ and $\dimension_y$, respectively.


\newcommand{\origspect}{s}
\newcommand{\recospect}{\tilde{s}}
\newcommand{\origwav}{w}
\newcommand{\recowav}{\tilde{w}}

\subsection{Audio Autoencoder}\label{subsec:autoencoder}
Our goal is to design an efficient audio autoencoder that can reach high compression ratios while reconstructing samples with reasonable accuracy.
To achieve this, we start from the audio autoencoder architecture proposed in Musika~\cite{musika}, where a model is used to reconstruct the magnitude and phase components of a spectrogram $s$ instead of the full waveform, which results in faster inference.
However, instead of using the original two-stage design and two-phase training process, we train a single encoder and decoder in a fully end-to-end fashion.
We first use a L1 loss between a log-magnitude spectrogram $s$ and the magnitude output of the model:
\begin{equation*}
\mathcal{L}_{E,D,\rec} = \mathbb{E}_{s \sim p(s)}||D(E(s))_{\magnitude}-s||_1
\end{equation*}
where $E$ and $D$ are the encoder and decoder, and $D(E(s))_{\magnitude}$ is the magnitude component of the decoder output.
We also use the multi-scale spectral distance~\cite{ddsp,rave} between the original and the reconstructed waveforms:
\begin{align*}
\recowav& =&\iSTFT(D(E(\origspect))) \\
  \mathcal{L}_{D,\mssd} &= &\mathbb{E}_{\origwav \sim p(\origwav)}\sum_{h\in \mathcal{H}} || \, \STFT_{h}(\origwav)^2 - \STFT_{h}(\recowav)^2 \, ||_1
\end{align*}
where $\mathcal{H}$ is a set of pairs of hop size and window length.
The phase component is modelled implicitly by the multi-scale spectral distance loss and the adversarial loss on the log-magnitude spectrogram of the reconstructed waveform:
\begin{align*}
\recospect& = &\log(\STFT(\recowav)^2+ \epsilon )\\
  \mathcal{L}_{C}& = &-\mathbb{E}_{\origspect \sim p(\origspect)}\left[ \min(0,\ -1+C(\origspect))\right] \\
          & &-\mathbb{E}_{\origspect \sim p(\origspect)}\left[\min(0,\ -1-C(\recospect))\right]\\
\mathcal{L}_{E,D,\adv}& = &-\mathbb{E}_{\origspect \sim p(\origspect)}\ C(\recospect)
\end{align*}
where $C$ is the critic.
The final objective used to jointly train encoder and decoder is the following:
\begin{align*}
\mathcal{L}_{E,D}&=\mathcal{L}_{E,D,\adv} + \lambdarec \mathcal{L}_{E,D,\rec} + \lambdamssd \mathcal{L}_{E,D,\mssd}
\end{align*}
Differently from~\cite{musika}, we add a second critic that receives mel-spectrograms.
This addition encourages the autoencoder to reconstruct spectral information more accurately in the regions where human pitch perception is more precise.


\begin{figure}[t!]
\centering
\includegraphics[width=0.5\textwidth]{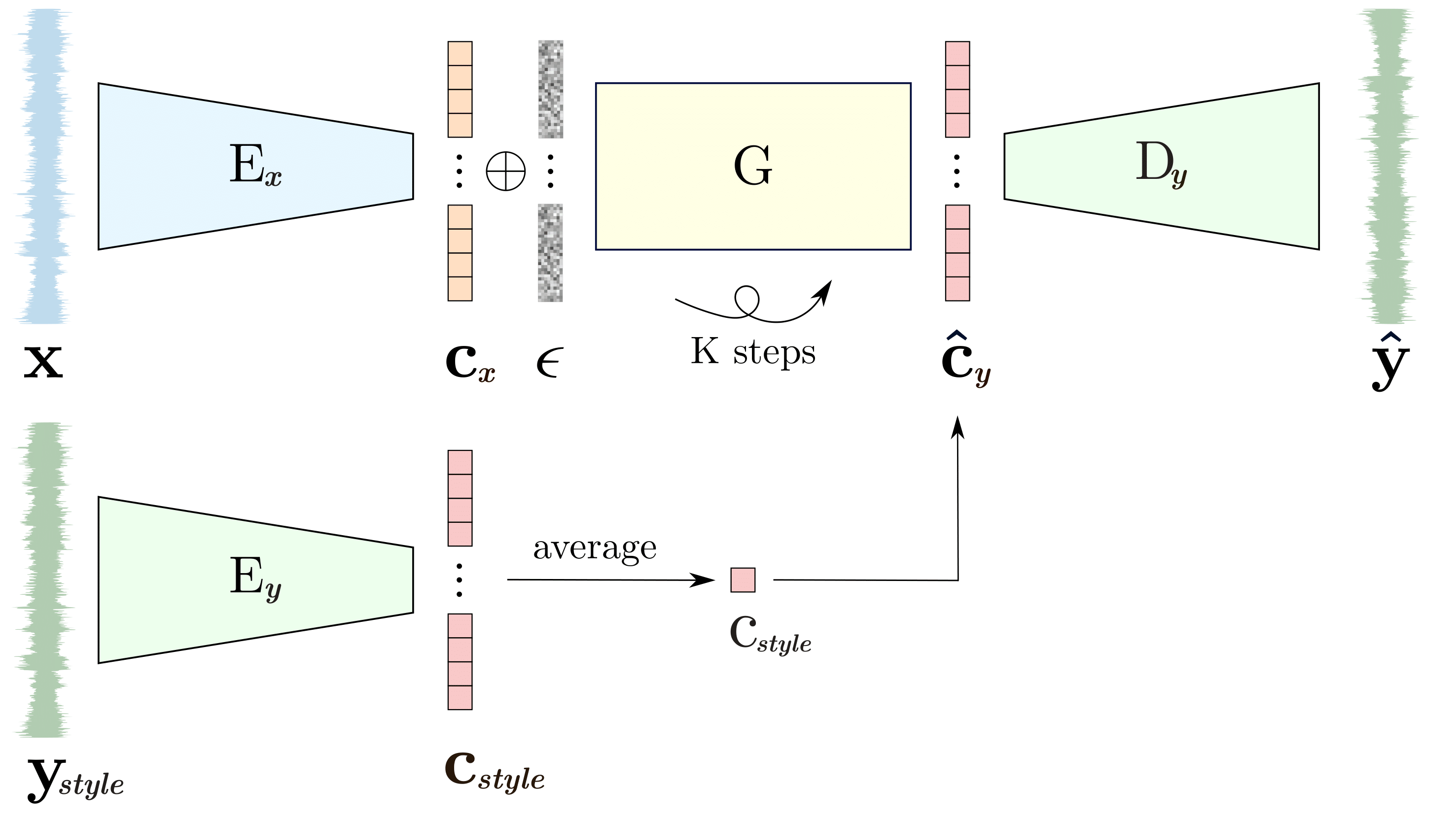}
\caption{Inference of the system.
Noise is concatenated to the latent representation of the conditioning waveform $\mathbf{c_x}$, and $K$ denoising steps are performed to generate $\mathbf{\hat{c}_y}$ which is then decoded to waveform.
The representation of a user-specified style sample $\mathbf{c_{style}}$ can be used to ground the generated output to a specific style.}
\label{fig:arch}
\end{figure}

\subsection{Latent Diffusion Model}
Diffusion models are trained to reverse a sequential corruption process of samples, and thus are able to retrieve samples from the data distribution by starting from a known distribution and iteratively denoising it.
We choose to briefly introduce them with their score-based interpretation~\cite{score}.

Our goal is to model the score of the conditional target stem latent distribution, given the input mix latent:
\begin{equation*}
G_{\theta}(\mathbf{c_y}, \mathbf{c_x}) \approx \nabla_{\mathbf{c_y}} \log p(\mathbf{c_y|c_x})
\end{equation*}
where $G_{\theta}(\mathbf{c_y}, \mathbf{c_x})$ is a neural network with parameters $\theta$.


To achieve this, we minimize the Fisher Divergence between the output of the model and score:
\begin{equation*}
\mathbb{E}_{p(\mathbf{c_y}, \mathbf{c_x})}\left[\left\|G_{\mathbf{\theta}}(\mathbf{c_y}, \mathbf{c_x})-\nabla_{\mathbf{c_y}} \log p(\mathbf{c_y|c_x})\right\|_2^2\right]
\end{equation*}
Finally, we can use Langevin dynamics to iteratively generate real samples with a sufficiently large number of iterations $K$.

In practice, we train our model to denoise noisy latent samples of the target stem $\mathbf{z}_t = \alpha_t \mathbf{c_y} +\beta_t \mathbf{\epsilon}$, with $\epsilon \sim \mathcal{N} (\mathbf{0},\mathbf{I})$:
\begin{equation*}
\mathcal{L}_{G_{\theta}} = \mathbb{E}_{\mathbf{c_y},\mathbf{c_x}\sim p(\mathbf{c_y}, \mathbf{c_x}),t\sim [0,1]}w_t||G_{\theta}(\mathbf{z}_t,t,\mathbf{c_x})-\mathbf{c_y}||_2^2
\end{equation*}
where $\alpha_t$ and $\beta_t$ are the signal and noise rates, $\mathbf{c}$ is the latent representation of the corresponding input mix and $w_t$ is the loss weight at timestep $t$.

The model is based on a U-Net architecture~\cite{unet}, with the addition of self-attention~\cite{transformer} in the lower resolution layers.
However, the vanilla self-attention mechanism does not allow the model to generalize to arbitrarily long inputs and outputs~\cite{extrapolate}, which is crucial for a flexible real-world use of the system.
To achieve generalization to lengths that are unseen during training, we equip the attention layers with Dynamic Positional Bias (DPB), a technique introduced for the task of arbitrarily-sized image classification~\cite{crossformer,swin2} which consists in the addition of a learnable Relative Positional Bias (RPB) matrix $\mathbf{B}\in \mathbb{R}^{L\times L}$ where $L$ is the temporal length of the feature map:
\begin{equation*}
\text{Attention}(\mathbf{Q},\mathbf{K},\mathbf{V})= \text{SoftMax}\left(\frac{\mathbf{QK^T}}{\sqrt{d}} + \mathbf{B}\right)
\end{equation*}
where $\mathbf{Q},\mathbf{K},\mathbf{V} \in \mathbb{R}^{L\times d}$ are query, key and value matrices.
Each entry $\mathbf{B}_{i,j}$ is learned with a Multi-Layer Perceptron (MLP) on the relative difference between positions $i$ and $j$:
\begin{equation*}
\mathbf{B}_{i,j}=\text{MLP}(i-j)
\end{equation*}

\subsection{Style Grounding}
To maximize its utility as a creative tool for music artists, our objective is a generation system that is controllable by the user.
To this end, we design a technique that enables the generation of single-stem samples with user-specified timbre characteristics and style.
Given a reference audio waveform $\mathbf{y}$ provided by the user to indicate their desired style, we first encode it to a compressed latent representation $\mathbf{c}_{\style}$ with the corresponding audio autoencoder.
Then, we simply average the latent representation over the timesteps to obtain a single $\dimension_y$ dimensional vector $\mu_{t}(\mathbf{c}_{\style})$, where $\mu_{t}(\cdot)$ indicates the average across all timesteps.
Finally, during the diffusion model sampling process, we force the generated latent samples at each reverse diffusion timestep to have an average across time that remains close to $\mu_{t}(\mathbf{c}_{\style})$.
We weigh this re-centering by the square of the timestep-specific noise rate, so that the effect is stronger at earlier iterations while keeping the model free to deviate when generating the lower-level details of the sample.
Given the denoised output of the diffusion model $\mathbf{\hat{c}}_{y, k} \in \mathbb{R}^{T \times \dimension_y}$ at sampling iteration $k$ we calculate:
\begin{equation*}
\mathbf{\hat{c}}_{y, k, \ground} = \mathbf{\hat{c}}_{y, k} - \mu_{t}(\mathbf{\hat{c}}_{y, k})  +  \beta_k^2 \mu_{t}(\mathbf{c}_{\style}) +(1-\beta_k^2) \mu_{t}(\mathbf{\hat{c}}_{y, k}) 
\end{equation*}

This technique exploits the semantically rich latent space produced by the autoencoder to enforce distinct timbre features captured in $\bar{\mathit{c}}_y$ onto the output of the diffusion model.

\subsection{Classifier-Free Guidance}
Classifier-Free Guidance (CFG)~\cite{cfg} is a technique that allows a conditional diffusion model to generate samples that more closely adhere to the provided input:
\begin{equation*}
\mathbf{\hat{c}}_{k, \cfg} = G_{\theta}(\mathbf{\hat{z}}_k,k,\mathbf{c_x}) +\lambda_{\cfg}(G_{\theta}(\mathbf{\hat{z}}_k,k)-G_{\theta}(\mathbf{\hat{z}}_k,k,\mathbf{c_x}))
\end{equation*}
where $G_{\theta}(\mathbf{\hat{z}}_k,k)$ is an unconditionally-generated sample at timestep $k$.
However, when high guidance weights $\lambda_{\cfg}$ are used, image generation models are known to generate overly saturated and exposed images~\cite{imagen}.
We experience a similar issue in our latent audio generation scenario, with highly distorted and saturated samples being generated.
Solutions such as clipping of the guided samples between a defined range of values or dynamic thresholding~\cite{imagen} are not applicable in our case, since our latent space is not bounded.
We thus use the technique proposed by~\cite{diffflaw} for guiding the generation of arbitrary spaces, which controls the increase in standard deviation of the guided samples with an hyperparameter $\phi \in [0,1]$, and allows us to reduce artifacts at higher guidance weights.
\begin{figure}[t]
\centering
\includegraphics[trim=30 5 5 5,width=.22\textwidth]{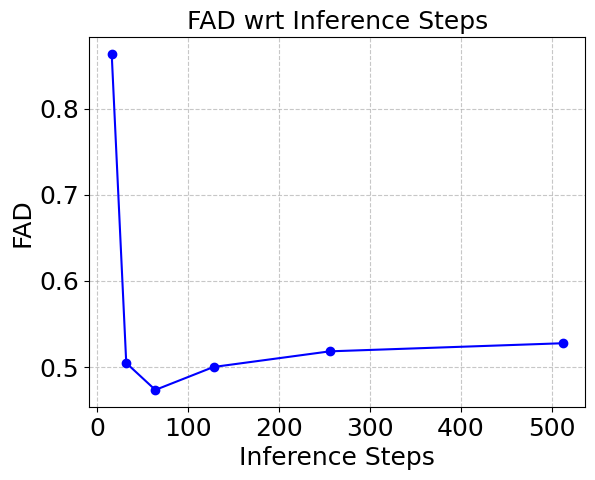}
\includegraphics[trim=5 5 20 5,width=.23\textwidth]{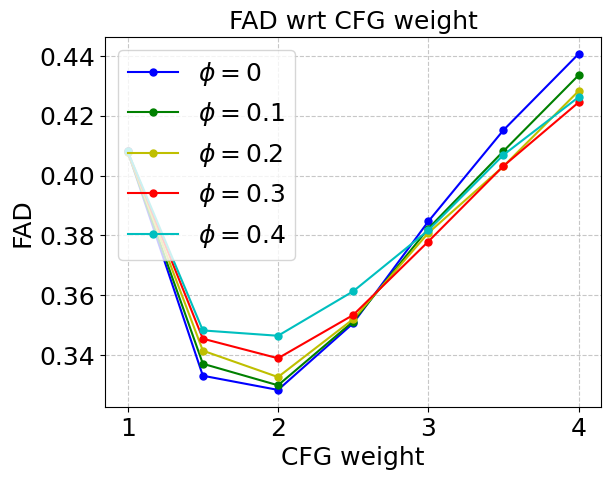}
\caption{\emph{Left}: FAD evaluation of unconditional samples with respect to the number of DDIM inference steps.
64 steps result in the lowest FAD, and we use $K=64$ in all subsequent experiments.
\emph{Right}: FAD evaluation of conditional samples with respect to CFG weights and with varying $\phi$.
When higher CFG weights ($>2.5$) are used, the latent rescaling technique results in lower FAD.}
\label{fig:uncondcond}
\end{figure}

\begin{figure}[t]
\centering
\includegraphics[clip,trim=40 5 20 5,width=0.4\textwidth]{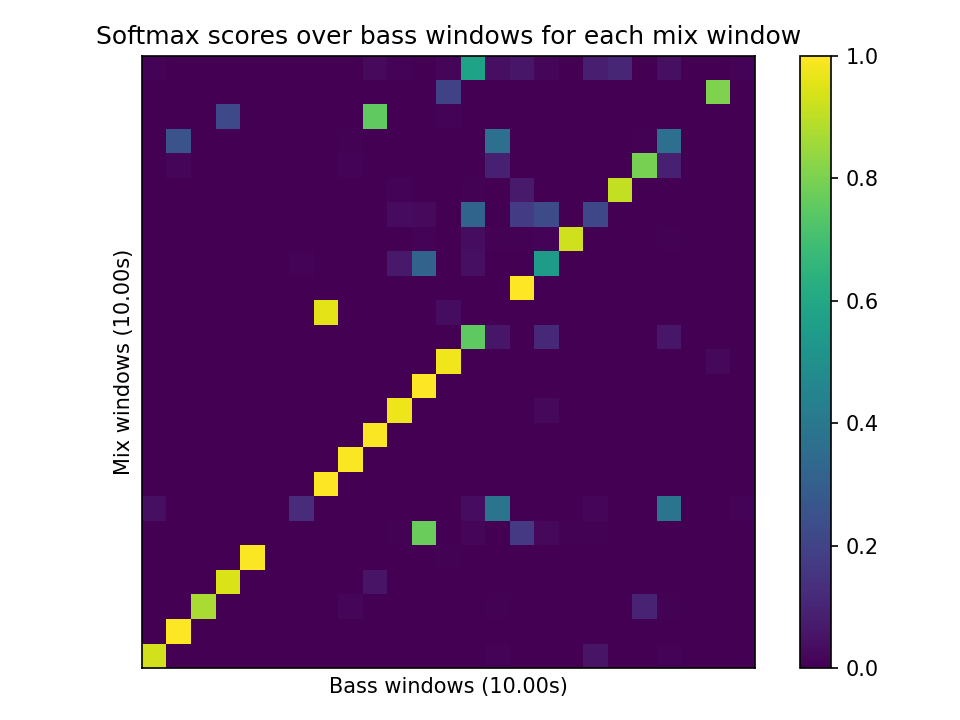}
\caption{Soft assignments of 25 random input mixes and corresponding generated basslines by a contrastive model (Section~\ref{sec:experiments}).
High diagonal values indicate the generated basslines best match their respective conditional inputs.}
\label{fig:softmax}
\end{figure}

\begin{table}[h!]
\footnotesize 
\centering
\begin{tabular}{l|l|l|}
\cline{2-3}
& Grounded & Not Grounded \\ \hline
\multicolumn{1}{|l|}{Cosine Distance}    & 0.269    & 0.644        \\ \hline
\multicolumn{1}{|l|}{Euclidean Distance} & 0.407    & 0.836        \\ \hline
\end{tabular}
\caption{Average Euclidean and Cosine distance between embeddings of style samples from the test set and embeddings of generated samples both using the proposed grounding technique and not using it.}
\label{table:ground}
\end{table}

\section{Implementation Details}
\label{sec:implementation_details}
We train the audio autoencoders on random crops of $1.5$ seconds to produce representations with $\dimension_x=64$ and $\dimension_y=32$, while $\rtime=4096$ is kept the same for both models.
Input log-magnitude spectrograms for both the autoencoder and the critics are calculated using $\hoplen=256$ and $\winlen=4 \cdot \hoplen$. $128$ mel-bins are used for the second critic.
The architecture of both autoencoder and critics consists of residual convolutional blocks.
We choose $\lambdarec=25, \lambdamssd=0.002$, and the multi-scale spectral distance loss is calculated using $\hoplen \in [2^5,2^6,2^7,2^8,2^9,2^{11},2^{12}]$. We always choose $\winlen = 4 \cdot \hoplen$.
The autoencoders consist of $37$M parameters and are trained using Adam~\cite{adam} with $\beta_1 = 0.5$ and $\beta_2 = 0.9$ for 500k iterations at a batch size of $32$.
The latent diffusion model is trained on (mix, stem) pairs, where both samples are $\sim$23 seconds long and are first encoded to 256 timesteps-long latent representations.
For a given track, the mix is obtained by mixing a non-empty random subset of stems from the track.
The latent diffusion model consists of residual convolutional blocks, with self-attention layers at the lower resolution levels.
The latent representation of the conditioning mix is concatenated with the noisy input, while the diffusion timestep information is expressed through sinusoidal embeddings~\cite{transformer} which are concatenated with the feature maps before every block.
$15\%$ of input latent representations are zero-ed out to train the model unconditionally, thus allowing CFG.
The latent diffusion model consists of $42$M parameters and is trained using AdamW~\cite{adamw} with $\beta_1 = 0.9$ and $\beta_2 = 0.999$ for 500k iterations at a batch size of $128$.
To train the model we use the v-objective \cite{progrdistill} with a cosine schedule, while at inference we use the DDIM sampler \cite{ddim}.



\section{Experiments and Results}
\label{sec:experiments}
We train the proposed accompaniment generation system on the task of conditional bassline generation, using an internal dataset of ~20\,k songs with available stems, among which the bass guitar.
1,500 of the tracks are used as test set.
We first train the audio autoencoder used to encode the input mixes on the MTG-Jamendo dataset~\cite{mtg}.
The autoencoder used to encode the bass samples is trained on bass stems from our internal dataset and the latent diffusion model is trained on (mix, bass stem) pairs from the same dataset.
We first evaluate the quality of unconditionally generated samples with respect to the number of DDIM steps in Fig.~\ref{fig:uncondcond} (right).
We show in Fig.~\ref{fig:uncondcond} (left) how the CFG rescaling technique can improve the FAD of generated samples for high CFG weights.
To evaluate the ability of the system to generate samples that musically match the input mix, we train a contrastive model to assign high scores to matching (mix, bass stem) pairs and low scores to non-matching ones using the same internal dataset. In Fig.~\ref{fig:softmax}, we visualize the scores assigned by that model to $25$ pairs of random segments of mixes from the test set, and 25 bass stems generated conditionally for each of those segments. A high value on the diagonal means the bass stem generated for that mix matches that mix better than the bass stems generated for the other mixes.
To quantitatively evaluate the efficacy of the proposed style grounding technique, we use an off-the-shelf audio classification model~\cite{pann} to extract embeddings of generated samples with and witout style-grounding (using the same input mix as conditioning), and compare them in Table \ref{table:ground} to embeddings of the target style sample via the Cosine and Euclidean distance. Readers can listen to samples generated by our system at: {\footnotesize{\url{https://sonycslparis.github.io/bass_accompaniment_demo/}}}

\vspace{-1.7ex}
\section{Conclusion}
\vspace{-.3ex}
\label{sec:conclusion}
We have presented a novel controllable system for music accompaniment generation using latent diffusion models.
When trained on bass stems, our model is able to generate basslines that musically match an arbitrary input mix.
We propose the design of an efficient audio autoencoder for producing compressed invertible latent representations, the adaptation of latent diffusion models to handle inputs and outputs of arbitrary length, and a latent-specific style grounding technique to control the timbre of generated samples.
Experiments demonstrate that our model can generate basslines that musically match the input mix and that can be grounded with user-provided timbres.
A limitation of our system is that it does not offer user control over the exact notes of the generated accompaniment.
Future work involves training the model to generate other instruments besides bass.
We believe our system can enhance the creative workflow of music artists, creating a variety of bass accompaniments to fit their existing material, while also offering control over the creation process.

\noindent\raisebox{-1ex}{\textit{This work was supported by UKRI
[grant EP/S022694/1].}}





\vfill\pagebreak




\printbibliography

\end{document}